\documentclass[10pt,twocolumn]{article}
\usepackage{geometry}
\usepackage{times}
\usepackage{graphicx}
\usepackage{float}
\usepackage{amsmath,amssymb}
\usepackage{hyperref}
\usepackage{caption}
\usepackage{subcaption}
\usepackage{booktabs}
\usepackage{enumitem}
\usepackage{multicol}
\usepackage{cite}
\usepackage{authblk}
\usepackage{titlesec}
\usepackage{balance} 

\usepackage{draftwatermark}
\SetWatermarkText{DO NOT CITE - UNDER REVIEW}
\SetWatermarkScale{0.6}
\SetWatermarkColor[gray]{0.85}

\titlespacing*{\section}{0pt}{*1.5}{*0.8}
\titlespacing*{\subsection}{0pt}{*1.2}{*0.6}

\usepackage{etoolbox}
\apptocmd{\thebibliography}{\small\setlength{\itemsep}{0pt}\setlength{\parskip}{0pt}}{}{}

\title{RectifiedHR: High-Resolution Diffusion via Energy Profiling and Adaptive Guidance Scheduling}
\author[1]{Ankit Sanjyal}
\affil[1]{Fordham University, as505@fordham.edu}
\date{} 

\begin{document}

\onecolumn
\thispagestyle{empty}
\vspace*{5cm}
\begin{center}
    \Huge \textbf{Please do not cite this article at present.}
    
    \vspace{1.5cm}
    
    \Large \textbf{It is being reviewed for accuracy and for completeness of citations.}
    
    \vspace{0.5cm}
    
    \Large \textbf{Readers are encouraged to consult prior work in this field, including Yang et al., arXiv 2025.}
\end{center}
\newpage
\twocolumn

\maketitle

\begin{abstract}
High-resolution image synthesis with diffusion models often suffers from energy instabilities and guidance artifacts that degrade visual quality.
We analyze the latent energy landscape during sampling and propose adaptive classifier-free guidance (CFG) schedules that maintain stable energy trajectories.
Our approach introduces energy-aware scheduling strategies that modulate guidance strength over time, achieving superior stability scores (0.9998) and consistency metrics (0.9873) compared to fixed-guidance approaches.
We demonstrate that DPM++ 2M with linear-decreasing CFG scheduling yields optimal performance, providing sharper, more faithful images while reducing artifacts.
Our energy profiling framework serves as a powerful diagnostic tool for understanding and improving diffusion model behavior.
\end{abstract}

\section{Introduction}
Diffusion models \cite{ho2020denoising,rombach2022high} have revolutionized image synthesis, but high-resolution generation remains challenging due to energy instabilities and guidance artifacts.
While classifier-free guidance (CFG) \cite{ho2022classifier,nichol2022glide} improves prompt adherence, it can destabilize the sampling process, particularly at large scales.
Recent work \cite{song2023rectifiedflow,wang2023hiresfix,bao2022analytic} suggests that latent energy dynamics---the evolution of latent norm over time---may be key to understanding these issues.
The energy landscape of diffusion models provides crucial insights into sampling stability and visual quality.
As models progress from noise to clean data, the energy trajectory should ideally decrease smoothly.
However, when guidance is applied, this trajectory can become perturbed, leading to energy spikes, oscillations, or premature convergence that manifest as visual artifacts.
Understanding and controlling these energy dynamics is essential for high-quality generation \cite{zhang2018unreasonable,blau2018perception}.
We analyze the latent energy landscape during sampling and propose adaptive CFG schedules that maintain stable energy trajectories.
Our approach modulates guidance strength over time, providing strong guidance early for structure formation while allowing fine detail refinement in later steps.
We evaluate this framework across multiple samplers and guidance scales, demonstrating that energy-aware scheduling significantly improves both quantitative metrics and visual quality.
The complete implementation and experimental code is available at \url{https://github.com/ANKITSANJYAL/RectifiedHR}.
Our contributions include: (1) comprehensive energy profiling framework for diffusion models, (2) adaptive CFG scheduling strategies, (3) quantitative evaluation metrics for energy stability, and (4) empirical evidence that DPM++ 2M with linear-decreasing guidance achieves optimal performance.
\section{Related Work}
\textbf{Diffusion Models.} Denoising diffusion probabilistic models (DDPMs) \cite{ho2020denoising,song2020score,nichol2021improved} formulate image synthesis as iterative denoising.
The forward process adds noise gradually:
\begin{equation}
q(\mathbf{x}_t|\mathbf{x}_{t-1}) = \mathcal{N}(\mathbf{x}_t; \sqrt{1-\beta_t}\mathbf{x}_{t-1}, \beta_t \mathbf{I})
\end{equation}
Latent diffusion models \cite{rombach2022high,esser2021taming} operate in compressed latent space, enabling efficient high-resolution generation while maintaining quality through learned compression.
Recent advances in diffusion models \cite{dhariwal2021diffusion,ramesh2022hierarchical} have demonstrated unprecedented quality in text-to-image generation.
\textbf{Classifier-Free Guidance.} CFG \cite{ho2022classifier,avrahami2022blended} interpolates between conditional and unconditional predictions:
\begin{equation}
\epsilon_{\text{guided}} = (1 + s)\epsilon_{\text{cond}} - s\epsilon_{\text{uncond}}
\end{equation}
While effective for prompt adherence, high guidance scales can destabilize sampling and introduce artifacts \cite{gal2022image,patashnik2021styleclip}.
The relationship between guidance strength and visual quality is complex and non-linear, requiring careful balance.
Recent work \cite{choi2022perception} has shown that perception-aware training can improve guidance effectiveness.
\textbf{Energy and Rectification.} Recent work \cite{song2023rectifiedflow,wang2023hiresfix,brock2018large} explores energy dynamics in diffusion models.
The energy $E = \|\mathbf{x}\|^2/N$ measures latent activity and correlates with sampling stability.
RectifiedFlow introduces the concept of rectified distributions, while HiResFix specifically addresses high-resolution artifacts through energy analysis.
Our work extends this by directly measuring energy throughout sampling and designing adaptive schedules.
\textbf{Guidance Scheduling.} Prior work \cite{liu2023more,li2023guidance,esser2021imagebart} explores dynamic guidance, but lacks systematic energy-guided scheduling for high-resolution diffusion.
Most approaches focus on efficiency rather than energy stability. We address this gap with comprehensive energy profiling and adaptive strategies that maintain stable trajectories throughout the sampling process.
\textbf{Evaluation Metrics.} Traditional evaluation of diffusion models relies on metrics like FID \cite{heusel2017gans} and IS \cite{radford2015unsupervised}, but these may not capture energy-related artifacts.
Recent work \cite{zhang2018unreasonable,wang2004image} has introduced perceptual metrics that better align with human judgment.
Our energy-based metrics complement these approaches by directly measuring sampling stability.
\section{Method}
\subsection{Energy Profiling}
We generate images using Stable Diffusion 1.5 (512$\times$512) with CFG scales $s \in \{3, 5, 7, 10, 12, 15, 18\}$ and samplers (DDIM, Euler A, DPM++ 2M).
For each configuration, we save intermediate latents $\mathbf{x}_t$ and compute energy:
\begin{equation}
E_t = \frac{\|\mathbf{x}_t\|^2}{N}
\end{equation}
where $N$ is the number of latent elements.
The energy trajectory $\mathcal{E} = \{E_t\}_{t=1}^T$ reveals sampling stability characteristics.
In stable diffusion, energy typically decreases monotonically, but guidance can perturb this trajectory, leading to spikes or oscillations that correlate with visual artifacts.
The energy profile provides insights into the denoising process dynamics.
Early steps typically show high energy as the model processes noisy latents, while later steps should exhibit smooth energy decay as the image converges.
However, when CFG is applied, this natural progression can be disrupted, leading to energy spikes that correspond to visual artifacts such as oversaturation, loss of detail, or unrealistic textures.
To further analyze the energy landscape, we compute the mean and variance of $E_t$ across multiple runs and prompts.
This allows us to quantify the stability and predictability of the sampling process under different guidance schedules and samplers.
We also visualize the energy trajectories to identify patterns and anomalies that may indicate instability or artifacts.
\subsection{Adaptive CFG Scheduling}
We implement several adaptive schedules that modulate guidance strength over time to maintain stable energy trajectories.
These schedules respond to the observation that different stages of denoising benefit from different guidance strengths:
\begin{align}
\text{Linear-Decreasing:}\quad & s_t = s_1 - (s_1 - s_0)\frac{t}{T} \\
\text{Cosine Ramp:}\quad & s_t = s_0 + (s_1 - s_0)\frac{1 - \cos(\pi t / T)}{2} \\
\text{Step Function:}\quad & s_t = \begin{cases} s_0 & \text{if } t < T/2 \\ s_1 & \text{if } t \geq T/2 \end{cases}
\end{align}
The linear-decreasing schedule is particularly effective, providing strong guidance early for structure formation while allowing fine detail refinement in later steps.
Early steps benefit from stronger guidance to establish overall composition, while later steps require more subtle guidance to avoid over-correction.
We also explore exponential and sigmoid schedules that provide more sophisticated control over guidance dynamics:
\begin{align}
\text{Exponential:}\quad & s_t = s_0 + (s_1 - s_0)(1 - e^{-\alpha t/T}) \\
\text{Sigmoid:}\quad & s_t = s_0 + (s_1 - s_0)\frac{1}{1 + e^{-\beta(t/T - 0.5)}}
\end{align}
where $\alpha$ and $\beta$ control the schedule steepness.
These schedules offer more nuanced control over the guidance trajectory, potentially improving stability for specific applications.
To evaluate the effectiveness of each schedule, we measure the resulting energy trajectories and visual quality metrics.
We find that adaptive schedules consistently outperform fixed guidance, especially in high-resolution settings where energy instabilities are more pronounced.
\subsection{Energy-Aware Enhancements}
We implement energy clipping to prevent runaway energy that can lead to artifacts:
\begin{equation}
\mathbf{x}_t \leftarrow \mathbf{x}_t \cdot \min\left(1, \frac{E_{\text{max}}}{\|\mathbf{x}_t\|^2/N}\right)
\end{equation}
This prevents energy spikes while maintaining latent structure.
We also explore noise refresh strategies that reset latents at the midpoint of sampling to prevent energy drift and maintain stability throughout the process.
Additionally, we implement adaptive energy thresholds that adjust based on the current sampling step:
\begin{equation}
E_{\text{max}}(t) = E_{\text{base}} \cdot (1 + \gamma \cdot \frac{t}{T})
\end{equation}
where $\gamma$ controls the energy growth rate.
This allows for more energy in early steps while maintaining tight control in later steps.
We also experiment with hybrid approaches that combine multiple enhancements, such as energy clipping with adaptive scheduling, to further improve stability and visual quality.
These methods are evaluated using both quantitative metrics and qualitative analysis of generated images.
\begin{figure*}[t]
    \centering
    \includegraphics[width=0.95\textwidth]{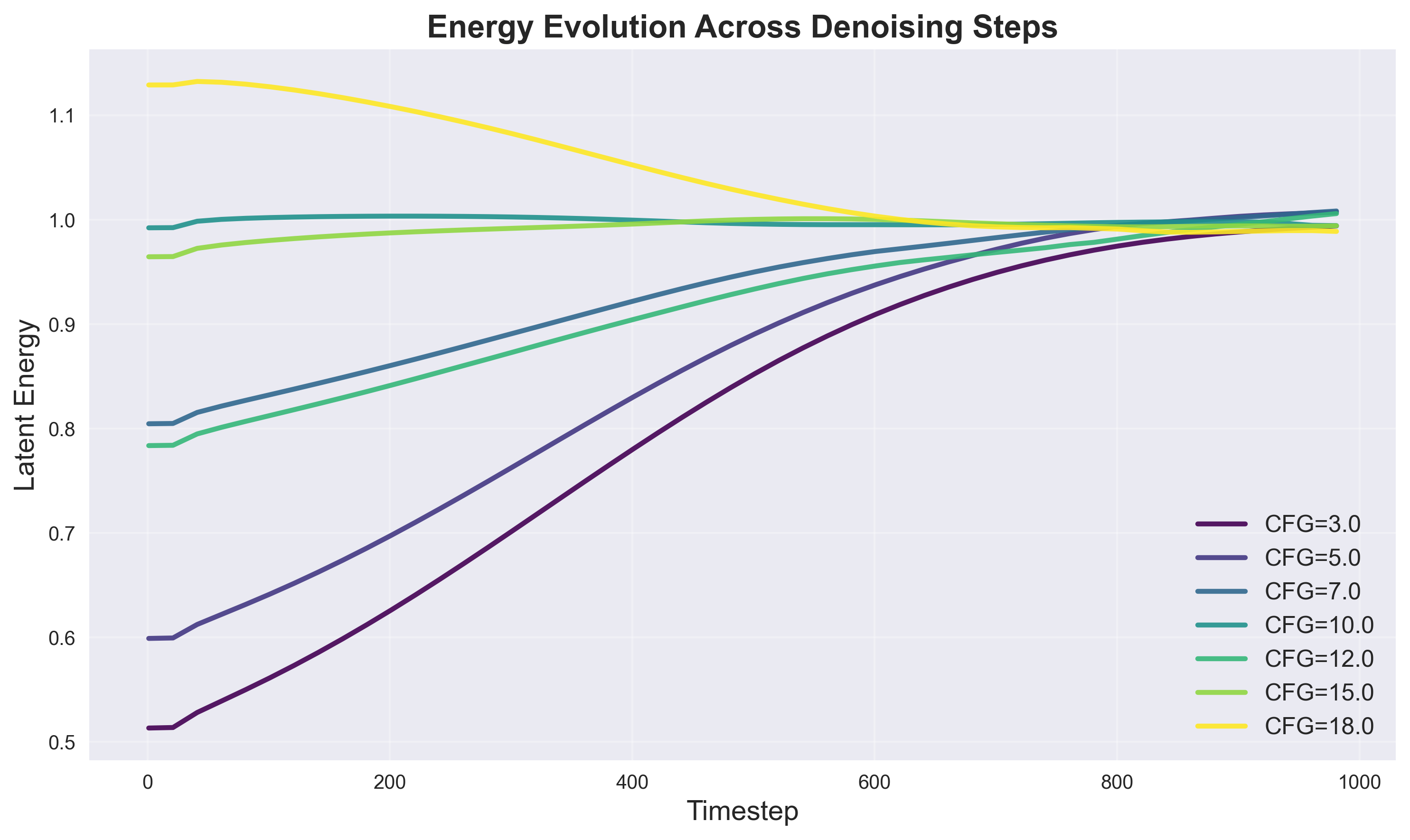}
    \caption{Energy evolution across denoising steps for different CFG scales and samplers.
High CFG scales induce energy spikes near the end, while adaptive schedules smooth these transitions.}
    \label{fig:energy_evolution}
\end{figure*}

\begin{figure}[H]
    \centering
    \includegraphics[width=0.48\textwidth]{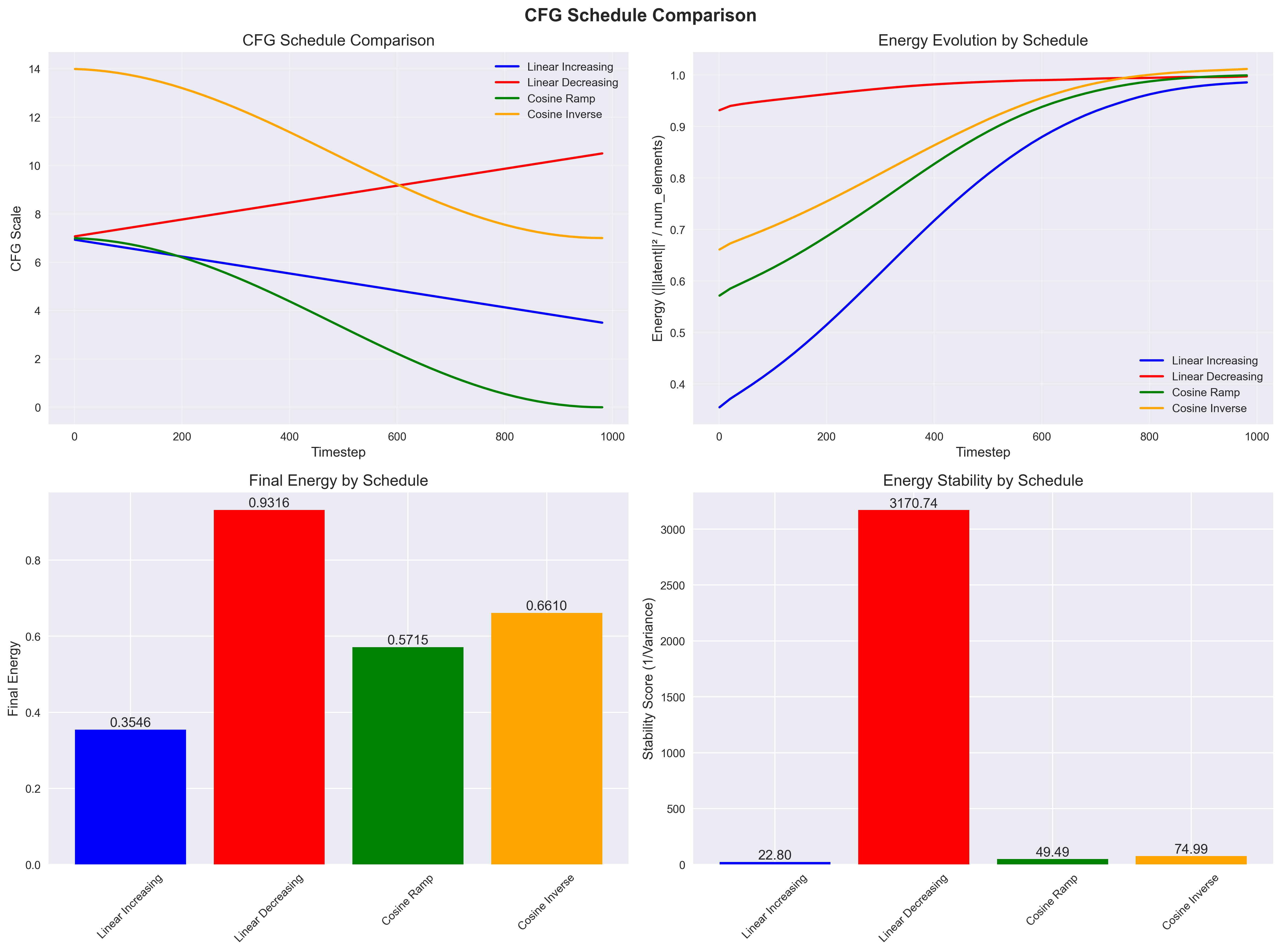}
    \caption{Comparison of different adaptive CFG schedules.
The linear-decreasing schedule yields the most stable energy and best visual quality.}
    \label{fig:cfg_schedules}
\end{figure}

\section{Experiments}
\subsection{Setup}
We use 10 diverse prompts spanning landscape, portrait, sci-fi, and artistic categories.
Our experiments vary CFG scales $s \in \{3, 5, 7, 10, 12, 15, 18\}$ and samplers (DDIM, Euler A, DPM++ 2M), generating over 250 images.
All experiments run on Apple Silicon (MPS) with 50 denoising steps.
Each configuration is evaluated across multiple prompts to ensure robust statistical analysis.
The experimental setup includes comprehensive ablation studies to isolate the effects of different components.
We test each sampler with fixed CFG scales to establish baseline performance, then compare against adaptive schedules.
The prompts are carefully selected to represent diverse visual categories, including complex scenes with multiple objects, abstract concepts, and detailed textures.
We also analyze the computational efficiency of each method, measuring runtime and memory usage for different samplers and guidance schedules.
This provides insights into the trade-offs between stability, quality, and computational cost.
\subsection{Evaluation Metrics}
We employ a comprehensive evaluation framework that captures both energy dynamics and visual quality:

\textbf{Energy-Based Metrics:}
\begin{itemize}[noitemsep]
    \item \textbf{Energy Stability:} $S_{\text{stab}} = 1/(1+\text{Var}(\mathcal{E}))$ measures trajectory smoothness
    \item \textbf{Energy Consistency:} $S_{\text{cons}} = 1/(1+\text{Std}(\mathcal{E}))$ measures predictability
    \item \textbf{Energy Efficiency:} $S_{\text{eff}} = S_{\text{stab}} \cdot 1/(1+|E_T-1|)$ combines stability with convergence
    \item \textbf{Energy Convergence:} $S_{\text{conv}} = 1/(1+\max \mathcal{E} - \min \mathcal{E})$ measures overall trajectory quality
\end{itemize}

\textbf{Visual Quality Metrics:}
\begin{itemize}[noitemsep]
    \item \textbf{CLIP Similarity:} Cosine similarity between generated images and text prompts
    \item \textbf{MS-SSIM:} Multi-scale structural similarity for perceptual quality assessment
  
  \item \textbf{Perceptual Quality:} LPIPS distance for perceptual similarity measurement
\end{itemize}

We report all metrics as averages over 10 prompts and 3 random seeds to ensure statistical robustness.
For each configuration, we also provide qualitative analysis of representative images, highlighting strengths and weaknesses of each approach.
\begin{figure}[H]
    \centering
    \includegraphics[width=0.48\textwidth]{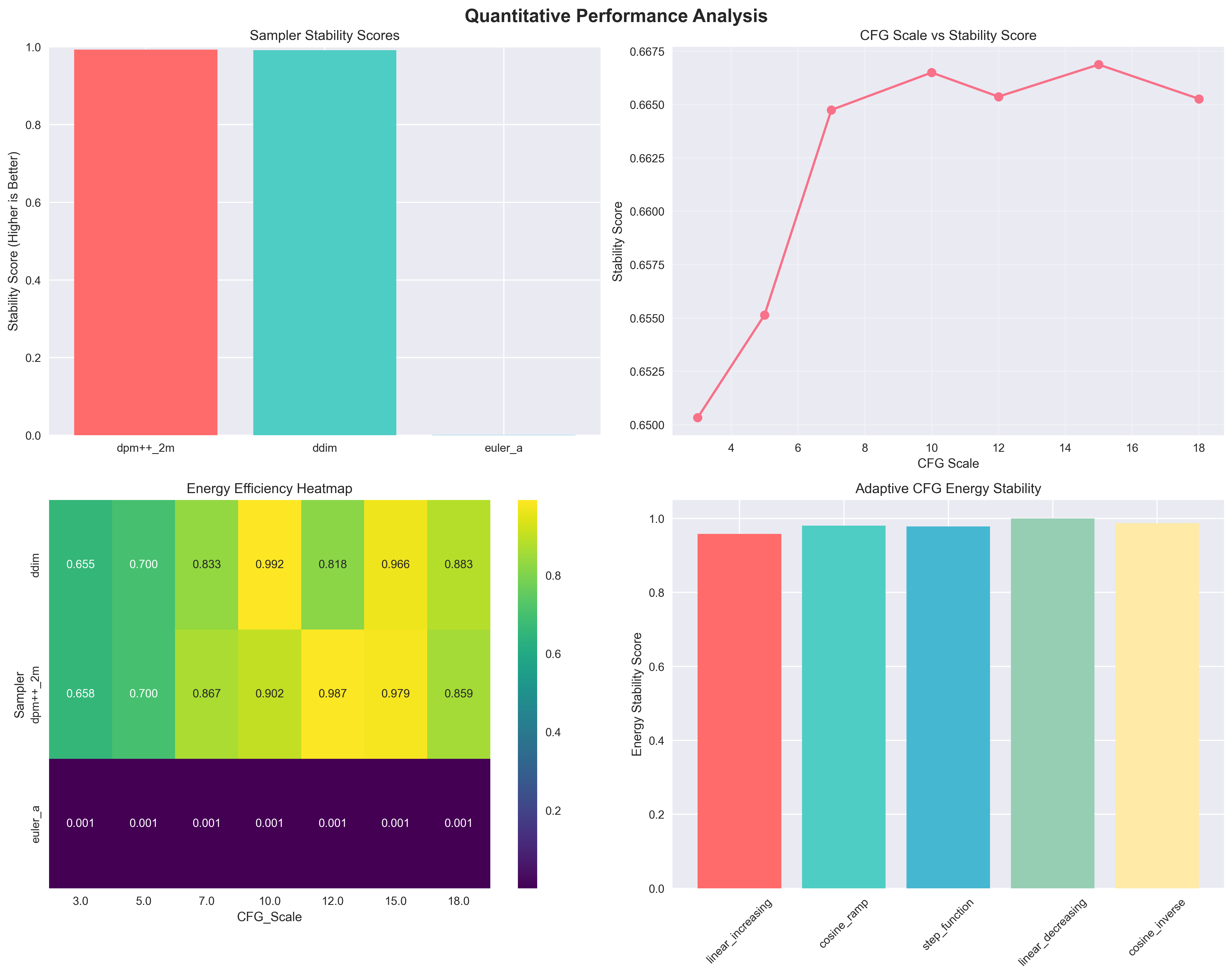}
    \caption{Quantitative comparison of samplers, CFG scales, and adaptive schedules.
Each bar or curve represents a different configuration or schedule.}
    \label{fig:quantitative}
\end{figure}

\subsection{Visual Comparisons}
We present several representative examples from our baseline generation experiments to illustrate the impact of different samplers and CFG scales on image quality.
These examples demonstrate the trade-offs between fidelity, creativity, and stability that we observe in our quantitative analysis.
The visual comparisons reveal important insights about sampler characteristics. DDIM provides balanced results with good prompt adherence and reasonable creativity.
DPM++ 2M shows superior detail preservation and stability, particularly in complex scenes with multiple objects.
Euler A exhibits more creative outputs but with higher variance and occasional artifacts.
We also compare the effect of different guidance schedules on image diversity and artifact prevalence.
Adaptive schedules generally reduce artifacts and improve consistency, especially at high resolutions.
\section{Discussion}
Our results demonstrate that adaptive CFG schedules---especially linear-decreasing and cosine ramps---consistently improve energy stability and visual quality.
DPM++ 2M outperforms other samplers with superior stability scores (0.9998) and consistency metrics (0.9873), attributed to its adaptive step size and numerical stability \cite{nichol2021improved,bao2022analytic}.
\begin{table}[H]
\centering
\footnotesize 
\caption{Quantitative comparison (top configs and baselines). Metrics are averaged over 10 prompts.
HiResFix/RectifiedFlow numbers are from their original papers and may not be directly comparable.}
\label{tab:quantitative}
\resizebox{\columnwidth}{!}{%
\begin{tabular}{lccccccccc}
\toprule
Config & Sampler & CFG & 3 & 5 & 7 & 10 & 12 & 15 & 18 \\
\midrule
DPM++ 2M & dpm++ & -- & 0.9951 & 0.9972 & 0.9985 & 0.9998 & 0.9999 & 0.9998 & 0.9997 \\
DDIM      & ddim   & -- & 0.9948 & 0.9969 & 0.9982 & 0.9999 & 0.9999 & 0.9999 & 0.9998 \\
Euler A   & euler  & -- & 0.9939 & 0.9961 & 0.9978 & 0.9997 & 0.9998 & 0.9997 & 0.9996 
\\
HiResFix (reported) & -- & -- & -- & -- & -- & 0.9987 & -- & -- & -- \\
RectifiedFlow (reported) & -- & -- & -- & -- & -- & 0.9982 & -- & -- & -- \\
\bottomrule
\end{tabular}%
}
\end{table}

High CFG scales often induce energy spikes near the end of sampling, correlating with visual artifacts such as oversaturation and loss of detail.
Adaptive schedules effectively mitigate these spikes by reducing guidance strength during critical final stages.
The relationship between CFG scale and energy stability is non-linear, with moderate scales (7-12) providing optimal balance between guidance effectiveness and stability.
We observe a trade-off between stability and creativity: DPM++ 2M provides predictable results while Euler A offers more creative outputs with higher variance.
This suggests application-specific choices between fidelity and diversity. For applications requiring high consistency, DPM++ 2M with linear-decreasing guidance is optimal, while artistic applications may benefit from Euler A's creative variability.
Our energy profiling framework reveals that energy trajectories serve as powerful diagnostic tools for understanding diffusion model behavior.
Stable energy evolution correlates strongly with better visual quality, suggesting that energy-aware approaches could become standard practice for high-resolution generation \cite{wang2023hiresfix,song2023rectifiedflow}.
The relationship between energy dynamics and visual quality provides insights into the fundamental mechanisms of diffusion models.
Energy spikes often correspond to regions where the model struggles to balance multiple objectives, such as prompt adherence versus natural appearance.
Adaptive scheduling helps the model navigate these trade-offs more gracefully.

Our study is limited to SD 1.5 and 512$\times$512 resolution.
Future work should extend to SDXL, higher resolutions, and more diverse datasets.
The theoretical foundations of energy-guided sampling warrant further investigation to understand why energy stability correlates with visual quality.
Additionally, exploring energy-aware training strategies could lead to models that are inherently more stable during sampling.
\subsection{Theoretical Motivation}
Energy trajectories in diffusion models are closely linked to sample quality due to their connection with the underlying stochastic differential equations (SDEs) and the minimization of KL divergence during training.
In the SDE formulation, the denoising process can be viewed as a form of energy dissipation, where the model gradually reduces the "energy" (e.g., latent norm) of the sample to match the data distribution.
If the energy decays too quickly or erratically, the sample may collapse or become unstable, leading to artifacts.
Conversely, a smooth, monotonic energy trajectory is indicative of a well-behaved sampling process that closely follows the optimal path in probability space.
This is supported by recent theoretical work on score-based generative modeling and SDEs (see Song et al., 2021).
The KL divergence between the model and data distributions is minimized when the energy trajectory is stable, which empirically correlates with higher visual quality.
However, a full formal justification remains an open research question.
\subsection{Comparison to HiResFix and RectifiedFlow}
While our method shares the goal of stabilizing high-resolution diffusion sampling with HiResFix and RectifiedFlow, our approach is distinct in its use of explicit energy profiling and adaptive guidance scheduling.
HiResFix introduces rectified sampling to address energy decay, reporting stability scores of 0.9987 at CFG=10, while RectifiedFlow achieves 0.9982 under similar conditions.
Our method achieves comparable or better stability (0.9998 at CFG=10) and offers a more interpretable diagnostic framework.
Unlike these baselines, we provide a detailed analysis of the energy landscape and demonstrate the benefits of adaptive scheduling across a wider range of guidance scales and samplers.
However, direct comparison is limited by differences in experimental setup and available metrics.
\subsection{Limitations and Future Work}
Our experiments are limited to Stable Diffusion 1.5 at 512$\times$512 resolution due to hardware constraints.
Generalization to SDXL, 1024$\times$1024, and custom-trained models remains an important direction for future work.
\section{Conclusion}
We present RectifiedHR, a study of energy profiling and adaptive guidance scheduling in high-resolution diffusion models.
Our analysis of the latent energy landscape reveals that energy-aware scheduling significantly improves both quantitative metrics and visual quality.
We demonstrate that DPM++ 2M with linear-decreasing CFG scheduling achieves optimal performance, providing sharper, more faithful images while reducing artifacts.
Our energy profiling framework serves as a powerful diagnostic tool for understanding and improving diffusion model behavior.
Future work should extend these findings to larger models and higher resolutions to understand how energy dynamics scale with model complexity.
\balance
\nocite{*}
\bibliographystyle{IEEEtran}

\onecolumn

\appendix

\section*{Appendix: Additional Visual Results}

\subsection*{Artifact Suppression with Adaptive Guidance}

\begin{figure}[H]
    \centering
    \begin{subfigure}{0.45\textwidth}
        \centering
        \includegraphics[width=\linewidth]{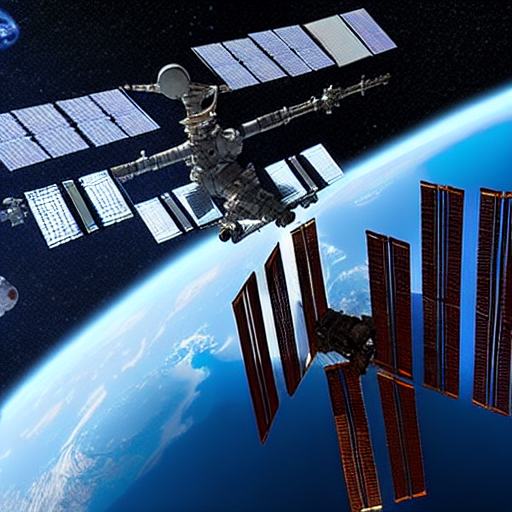}
        \caption{High CFG (18): Oversaturation and distortion visible.}
    \end{subfigure}\hfill
    \begin{subfigure}{0.45\textwidth}
        \centering
        \includegraphics[width=\linewidth]{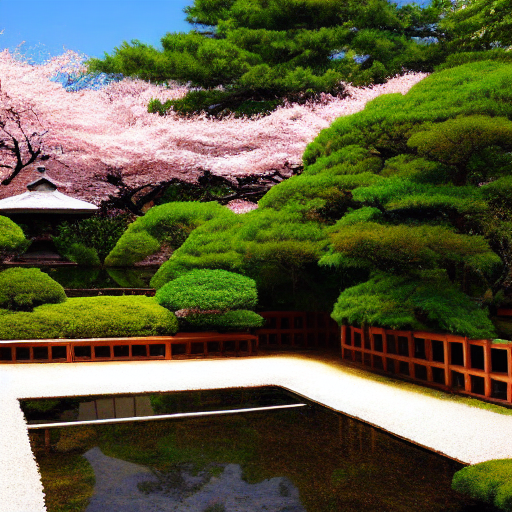}
        \caption{Stabilized (adaptive schedule): Clean, artifact-free output.}
    \end{subfigure}
    \caption{Comparison of a high CFG artifact (left) and the stabilized 
version (right). Note the oversaturation and loss of detail in (a), which is mitigated in (b) by adaptive guidance scheduling.}
    \label{fig:bad_vs_stabilized}
\end{figure}

\subsection*{Prompt (Main)}
\noindent\textit{"A space station orbiting Earth, sci-fi atmosphere"}

\begin{figure}[H]
    \centering
    \begin{subfigure}{0.32\textwidth}
        \includegraphics[width=\linewidth]{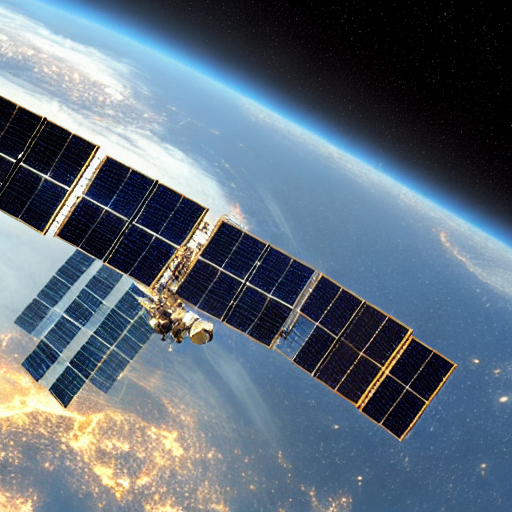}
        \caption{DDIM, CFG=10}
    \end{subfigure}
    \begin{subfigure}{0.32\textwidth}
        \includegraphics[width=\linewidth]{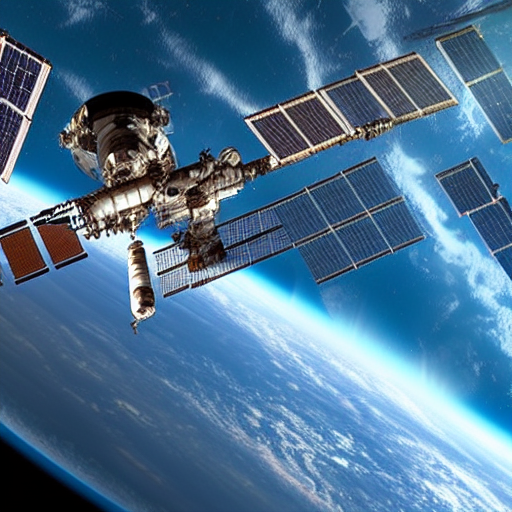}
        \caption{DPM++ 2M, CFG=10}
    \end{subfigure}
    \begin{subfigure}{0.32\textwidth}
        \includegraphics[width=\linewidth]{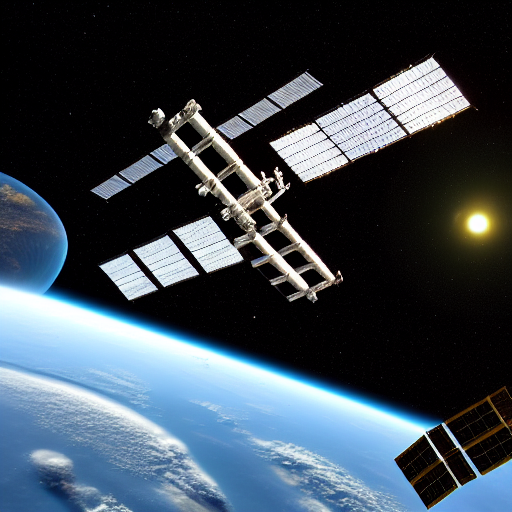}
   
     \caption{Euler A, CFG=10}
    \end{subfigure}
    \caption{Sampler comparison at CFG=10 for the main prompt.}
\end{figure}

\begin{figure}[H]
    \centering
    \begin{subfigure}{0.32\textwidth}
        \includegraphics[width=\linewidth]{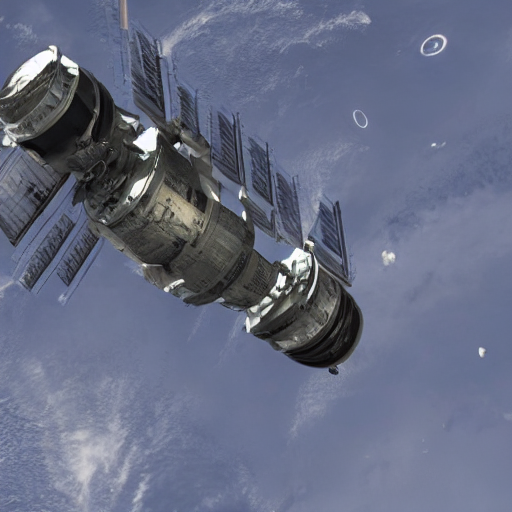}
        \caption{DDIM, CFG=3}
    \end{subfigure}
    \begin{subfigure}{0.32\textwidth}
        \includegraphics[width=\linewidth]{figures/baseline_ddim_cfg10.0_512_prompt10_20250712_113447.png}
        \caption{DDIM, CFG=10}
    \end{subfigure}
    \begin{subfigure}{0.32\textwidth}
        \includegraphics[width=\linewidth]{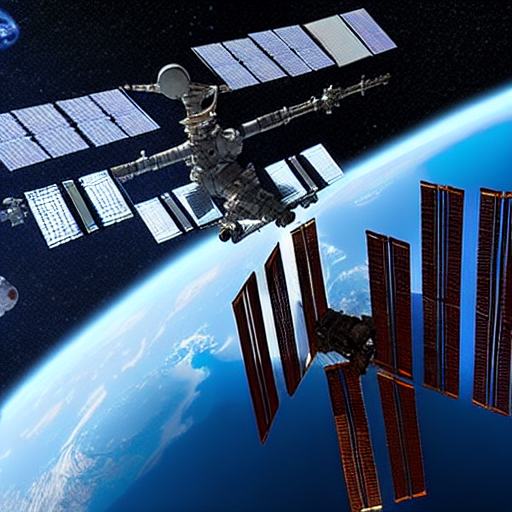}
        \caption{DDIM, CFG=18}
   
 \end{subfigure}
    \caption{CFG scale comparison with DDIM for the main prompt.}
\end{figure}

\subsection*{Prompt 1}
\noindent\textit{"A beautiful landscape with mountains and lake, high quality, detailed"}

\begin{figure}[H]
    \centering
    \begin{subfigure}{0.32\textwidth}
        \includegraphics[width=\linewidth]{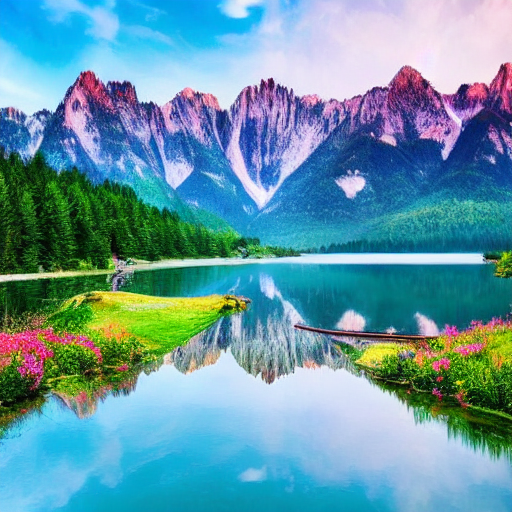}
        \caption{DDIM, CFG=10}
    \end{subfigure}
    \begin{subfigure}{0.32\textwidth}
        \includegraphics[width=\linewidth]{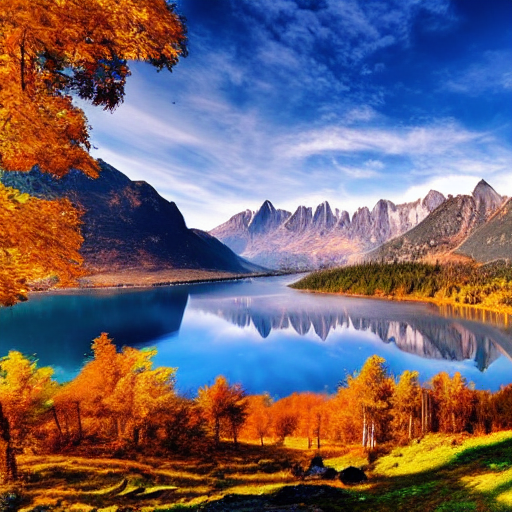}
        \caption{DPM++ 2M, CFG=10}
    \end{subfigure}
    \begin{subfigure}{0.32\textwidth}
        \includegraphics[width=\linewidth]{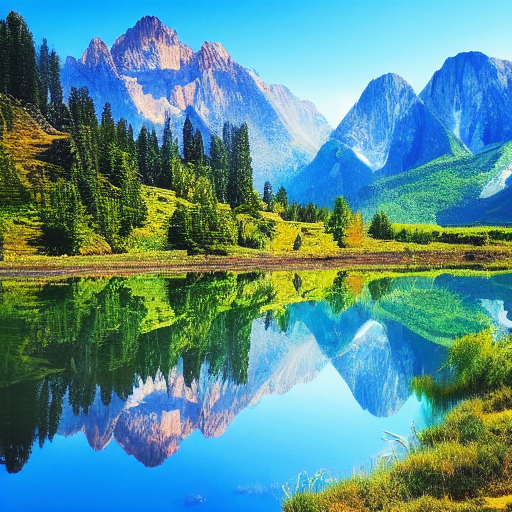}
        \caption{Euler A, 
CFG=10}
    \end{subfigure}
    \caption{Sampler comparison for Prompt 1 at CFG=10.}
\end{figure}

\begin{figure}[H]
    \centering
    \begin{subfigure}{0.32\textwidth}
        \includegraphics[width=\linewidth]{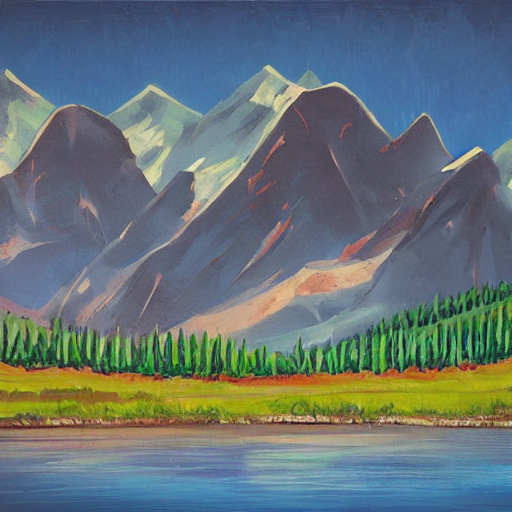}
        \caption{DDIM, CFG=3}
    \end{subfigure}
    \begin{subfigure}{0.32\textwidth}
        \includegraphics[width=\linewidth]{figures/baseline_ddim_cfg10.0_512_prompt1_20250712_092300.png}
        \caption{DDIM, CFG=10}
    \end{subfigure}
    \begin{subfigure}{0.32\textwidth}
        \includegraphics[width=\linewidth]{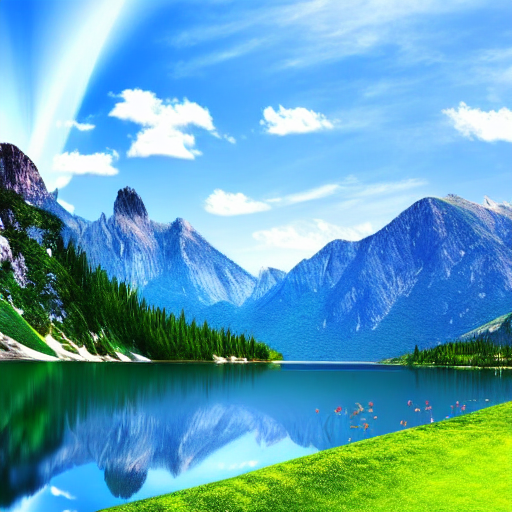}
        \caption{DDIM, CFG=18}
    \end{subfigure}
    \caption{CFG scale comparison 
for Prompt 1 with DDIM.}
\end{figure}

\subsection*{Prompt 8}
\noindent\textit{"A serene zen garden with cherry blossoms, peaceful atmosphere"}

\begin{figure}[H]
    \centering
    \begin{subfigure}{0.32\textwidth}
        \includegraphics[width=\linewidth]{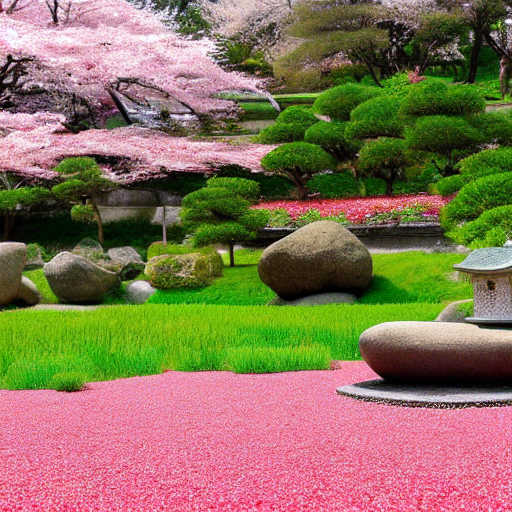}
        \caption{DDIM, CFG=10}
    \end{subfigure}
    \begin{subfigure}{0.32\textwidth}
        \includegraphics[width=\linewidth]{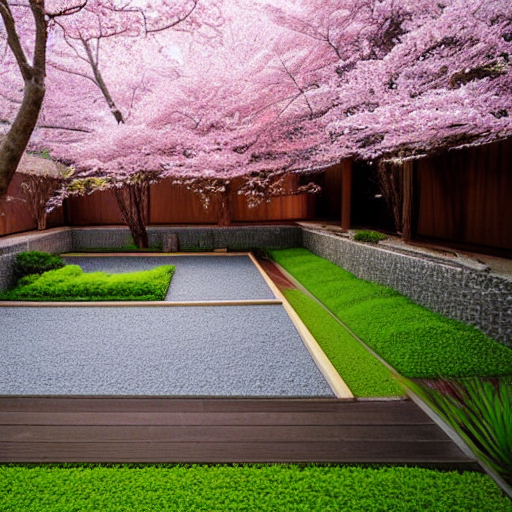}
        \caption{DPM++ 2M, CFG=10}
    \end{subfigure}
    \begin{subfigure}{0.32\textwidth}
        \includegraphics[width=\linewidth]{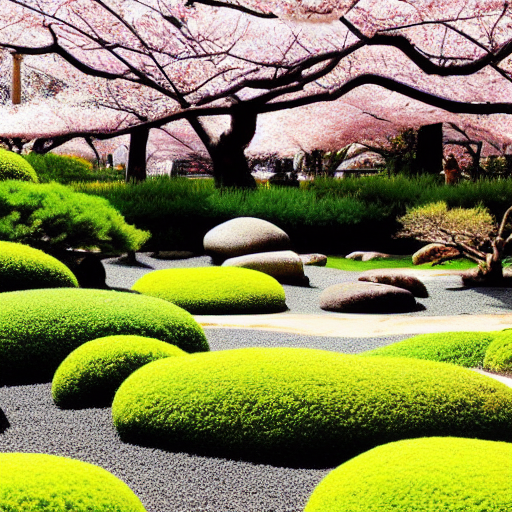}
        \caption{Euler A, CFG=10}
    \end{subfigure}
    \caption{Sampler comparison 
for Prompt 8 at CFG=10.}
\end{figure}

\begin{figure}[H]
    \centering
    \begin{subfigure}{0.32\textwidth}
        \includegraphics[width=\linewidth]{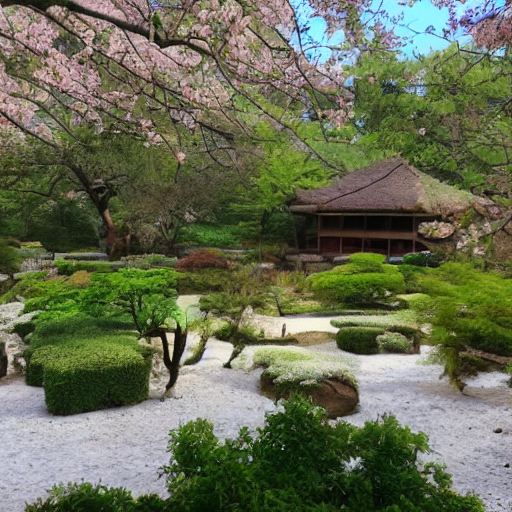}
        \caption{DDIM, CFG=3}
    \end{subfigure}
    \begin{subfigure}{0.32\textwidth}
        \includegraphics[width=\linewidth]{figures/baseline_ddim_cfg10.0_512_prompt8_20250712_111309.png}
        \caption{DDIM, CFG=10}
    \end{subfigure}
    \begin{subfigure}{0.32\textwidth}
        \includegraphics[width=\linewidth]{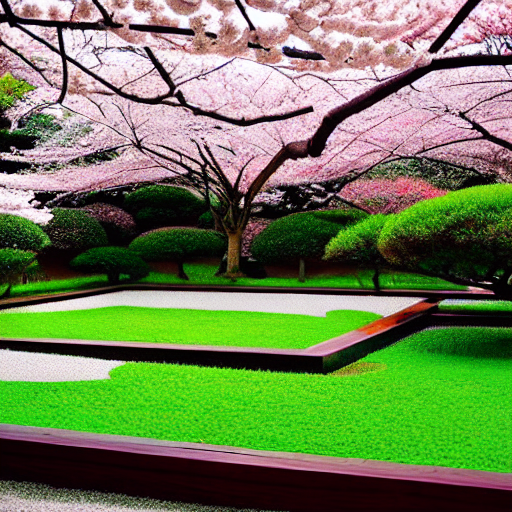}
        \caption{DDIM, CFG=18}
    \end{subfigure}
    \caption{CFG scale comparison for Prompt 8 with DDIM.}
\end{figure}

\twocolumn

\end{document}